\documentclass[prl,amsfonts,amssymb,showpacs,twocolumn]{revtex4}
\input epsf.sty

\def\met{\mbox{${\hbox{$E$\kern-0.6em\lower-.1ex\hbox{/}}}_T$}} 
\def\D0{D\O}                            

\def\ct{\mbox{$\cos\theta^*$}}

\begin{document}
\title{\vspace*{-0.25in}
Search for Large Extra Dimensions in Dielectron and Diphoton Production}
\vspace*{-0.5in}
%
\author{                                                                      
B.~Abbott,$^{50}$                                                             
M.~Abolins,$^{47}$                                                            
V.~Abramov,$^{23}$                                                            
B.S.~Acharya,$^{15}$                                                          
D.L.~Adams,$^{57}$                                                            
M.~Adams,$^{34}$                                                              
G.A.~Alves,$^{2}$                                                             
N.~Amos,$^{46}$                                                               
E.W.~Anderson,$^{39}$                                                         
M.M.~Baarmand,$^{52}$
% ...
\\                                                                            
\vskip 0.30cm                                                                 
\centerline{(D\O\ Collaboration)}                                             
\vskip 0.30cm                                                                 
}                                                                             
\address{                                                                     
\centerline{$^{1}$Universidad de Buenos Aires, Buenos Aires, Argentina}       
\centerline{$^{2}$LAFEX, Centro Brasileiro de Pesquisas F{\'\i}sicas,         
                  Rio de Janeiro, Brazil}                                     
\centerline{$^{3}$Universidade do Estado do Rio de Janeiro,                   
                  Rio de Janeiro, Brazil}                                     
\centerline{$^{4}$Institute of High Energy Physics, Beijing,                  
                  People's Republic of China}                                 
\centerline{$^{5}$Universidad de los Andes, Bogot\'{a}, Colombia}             
\centerline{$^{6}$Charles University, Prague, Czech Republic}                 
\centerline{$^{7}$Institute of Physics, Academy of Sciences, Prague,          
                  Czech Republic}                                             
\centerline{$^{8}$Universidad San Francisco de Quito, Quito, Ecuador}         
\centerline{$^{9}$Institut des Sciences Nucl\'eaires, IN2P3-CNRS,             
                  Universite de Grenoble 1, Grenoble, France}                 
\centerline{$^{10}$CPPM, IN2P3-CNRS, Universit\'e de la M\'editerran\'ee,     
                  Marseille, France}                                          
\centerline{$^{11}$LPNHE, Universit\'es Paris VI and VII, IN2P3-CNRS,         
                  Paris, France}                                              
\centerline{$^{12}$DAPNIA/Service de Physique des Particules, CEA, Saclay,    
                  France}                                                     
\centerline{$^{13}$Panjab University, Chandigarh, India}                      
\centerline{$^{14}$Delhi University, Delhi, India}                            
\centerline{$^{15}$Tata Institute of Fundamental Research, Mumbai, India}     
\centerline{$^{16}$Seoul National University, Seoul, Korea}                   
\centerline{$^{17}$CINVESTAV, Mexico City, Mexico}                            
\centerline{$^{18}$FOM-Institute NIKHEF and University of                     
                  Amsterdam/NIKHEF, Amsterdam, The Netherlands}               
\centerline{$^{19}$University of Nijmegen/NIKHEF, Nijmegen, The               
                  Netherlands}                                                
\centerline{$^{20}$Institute of Nuclear Physics, Krak\'ow, Poland}            
\centerline{$^{21}$Institute for Theoretical and Experimental Physics,        
                   Moscow, Russia}                                            
\centerline{$^{22}$Moscow State University, Moscow, Russia}                   
\centerline{$^{23}$Institute for High Energy Physics, Protvino, Russia}       
\centerline{$^{24}$Lancaster University, Lancaster, United Kingdom}           
\centerline{$^{25}$University of Arizona, Tucson, Arizona 85721}              
\centerline{$^{26}$Lawrence Berkeley National Laboratory and University of    
                  California, Berkeley, California 94720}                     
\centerline{$^{27}$University of California, Davis, California 95616}         
\centerline{$^{28}$California State University, Fresno, California 93740}     
\centerline{$^{29}$University of California, Irvine, California 92697}        
\centerline{$^{30}$University of California, Riverside, California 92521}     
\centerline{$^{31}$Florida State University, Tallahassee, Florida 32306}      
\centerline{$^{32}$University of Hawaii, Honolulu, Hawaii 96822}              
\centerline{$^{33}$Fermi National Accelerator Laboratory, Batavia,            
                   Illinois 60510}                                            
\centerline{$^{34}$University of Illinois at Chicago, Chicago,                
                   Illinois 60607}                                            
\centerline{$^{35}$Northern Illinois University, DeKalb, Illinois 60115}      
\centerline{$^{36}$Northwestern University, Evanston, Illinois 60208}         
\centerline{$^{37}$Indiana University, Bloomington, Indiana 47405}            
\centerline{$^{38}$University of Notre Dame, Notre Dame, Indiana 46556}       
\centerline{$^{39}$Iowa State University, Ames, Iowa 50011}                   
\centerline{$^{40}$University of Kansas, Lawrence, Kansas 66045}              
\centerline{$^{41}$Kansas State University, Manhattan, Kansas 66506}          
\centerline{$^{42}$Louisiana Tech University, Ruston, Louisiana 71272}        
\centerline{$^{43}$University of Maryland, College Park, Maryland 20742}      
\centerline{$^{44}$Boston University, Boston, Massachusetts 02215}            
\centerline{$^{45}$Northeastern University, Boston, Massachusetts 02115}      
\centerline{$^{46}$University of Michigan, Ann Arbor, Michigan 48109}         
\centerline{$^{47}$Michigan State University, East Lansing, Michigan 48824}   
\centerline{$^{48}$University of Nebraska, Lincoln, Nebraska 68588}           
\centerline{$^{49}$Columbia University, New York, New York 10027}             
\centerline{$^{50}$New York University, New York, New York 10003}             
\centerline{$^{51}$University of Rochester, Rochester, New York 14627}        
\centerline{$^{52}$State University of New York, Stony Brook,                 
                   New York 11794}                                            
\centerline{$^{53}$Brookhaven National Laboratory, Upton, New York 11973}     
\centerline{$^{54}$Langston University, Langston, Oklahoma 73050}             
\centerline{$^{55}$University of Oklahoma, Norman, Oklahoma 73019}            
\centerline{$^{56}$Brown University, Providence, Rhode Island 02912}          
\centerline{$^{57}$University of Texas, Arlington, Texas 76019}               
\centerline{$^{58}$Texas A\&M University, College Station, Texas 77843}       
\centerline{$^{59}$Rice University, Houston, Texas 77005}                     
\centerline{$^{60}$University of Washington, Seattle, Washington 98195}       
}                                                                             

\date{August 25, 2000}

\begin{abstract}
We report a search for effects of large extra spatial dimensions in $p{\bar p}$ collisions at a center-of-mass energy of 1.8 TeV with the D\O\ detector, using events containing a pair of electrons or photons. The data are in good agreement with the expected background and do not exhibit evidence for large extra dimensions. We set the most restrictive lower limits to date, at the 95\% confidence level, on the effective Planck scale between 1.0~TeV and 1.4~TeV for several formalisms and numbers of extra dimensions.
\end{abstract}

\maketitle

The possibility that the universe has more than three spatial dimensions has long been discussed~\cite{Riemann}. Recent developments in string theory suggest that there could be up to seven additional spatial dimensions, compactified at very small distances, on the order of $10^{-32}$~m. In a new model~\cite{ADD}, inspired by string theory, several of the compactified extra dimensions (ED) are suggested to be as large as 1~mm. These large ED are introduced to solve the hierarchy problem of the standard model (SM) by lowering the Planck scale, $M_{\rm Pl}$, to the TeV energy range. The ED compactification radius, $R$, depends on the number of extra dimensions ($n$) and on the effective Planck scale, $M_S$: $R \propto \frac{1}{M_S}(M_{\rm Pl}/M_S)^{2/n}$~\cite{ADD}. Since Newton's law of gravity would be modified in the presence of compactified extra dimensions for interaction distances below the size of the largest extra dimension, current gravitational observations rule out the case of a single large extra dimension. Recent preliminary results from gravity experiments at submillimeter distances~\cite{tabletop}, as well as cosmological constraints from supernova cooling and distortion of cosmic diffuse gamma-radiation~\cite{cosmology}, indicate that the case of $n = 2$ is likely ruled out as well. However, for $n \ge 3$, the size of the ED becomes microscopic and therefore eludes the reach of direct gravitational measurements. Cosmological constraints are also weak in this case. 
Therefore, high energy colliders, capable of probing very short distances, are crucial to test theories of large ED. In these theories, the effects of gravity are enhanced at high energies due to the accessibility of numerous excited states of the graviton (referred to as a Kaluza-Klein~\cite{KK} graviton, $G_{\rm KK}$), corresponding to multiple winding modes of the graviton field around the compactified dimensions. Since gravitons couple to the energy-momentum tensor, they can be produced in any SM process.

Large ED phenomenology at colliders has been studied in some detail~\cite{GRW,HLZ,Peskin,Hewett}. One of the primary observable effects would be an apparent non-conservation of momentum caused by the direct emission of  gravitons, which leave the three flat spatial dimensions. A typical signature would be the production of a single jet or vector boson at large transverse momentum. The other observable effect would be anomalous difermion or diboson production at large invariant masses ($M$) via virtual graviton exchange. Direct graviton emission is expected to be suppressed by a factor $(M/M_S)^{n+2}$, while virtual graviton effects depend only weakly on the number of extra dimensions~\cite{GRW,Hewett}. Virtual graviton production therefore offers a potentially more sensitive way to search for manifestations of ED~\cite{caveat}.

Both of the above effects have been sought at LEP~\cite{LEP}, with lower limits on the effective Planck scale set on the order of 1~TeV. Virtual $G_{\rm KK}$ exchange was also sought at HERA~\cite{HERA}, but with less stringent limits. In this Letter, we report the results of the first specific search for ED at a hadron collider. The data are from the D\O\ experiment~\cite{D0} at the Fermilab Tevatron, using proton-antiproton collisions at a center-of-mass energy of 1.8~TeV, and final states containing pairs of electrons or photons. We analyze the differential distribution~\cite{KCGL} of dielectron or diphoton events in terms of their invariant mass and the scattering angle in their center-of-mass frame ($\theta^*$). These two variables completely define the $2 \to 2$ scattering processes, and their properties provide improved separation between the SM contributions and the effects of virtual graviton exchange when compared to analyses of invariant-mass distributions alone~\cite{KCGL}.

Several modifications to the method of Ref.~\cite{KCGL} are introduced to optimize the sensitivity of the search. First, because the efficiency and resolution for high-energy electromagnetic (EM) objects at D\O\ are superior to those for muons, we use only the dielectron and diphoton channels, with the sensitivity to ED coming primarily from the diphoton events~\cite{KCGL}. Second, to optimize the efficiency for diphoton and dielectron selection, we eliminate tracking requirements from electron and photon identification, thereby effectively combining them. Ignoring tracking information maximizes both the  dielectron and diphoton efficiencies since neither electrons with unreconstructed tracks nor photons with matching tracks from conversion or random overlap are lost. In what follows, we refer to electron or photon objects that do not use tracking information as EM objects. The dominant background at high mass, where ED effects are enhanced, is the irreducible SM background from direct diphoton production, rather than instrumental background from misidentification of jets as EM objects (see, e.g., Ref.~\cite{GLKM}). Increasing the di-EM identification (ID) efficiency by using looser EM ID criteria therefore results in better sensitivity to ED.

The search is based on the entire sample of data collected during 1992--1996 using a trigger that requires the presence of two EM objects and corresponds to an integrated luminosity of $126.8 \pm 5.6$~pb$^{-1}$.

Following offline reconstruction, we require each EM object to: (i) deposit more than  95\% of its energy in the EM calorimeter; (ii) have an energy deposition pattern consistent with that expected for an EM object; and (iii) be isolated~\cite{Zvvg}. The efficiency for these selections, based on $Z \to ee$ events, is $87 \pm 2\%$ per EM object. We also require (i) two EM objects with $E_T >  45$~GeV in good fiducial regions of the detector ($|\eta_d|  < 1.1$ or $1.5 < |\eta_d| < 2.5$, where $\eta_d$ is the pseudorapidity relative to the center of the detector~\cite{eta}); (ii) missing transverse  energy $\met <  25$~GeV; and (iii) no additional EM objects with $E_T > 5$~GeV. The trigger is fully efficient for this set of offline selections. The above selections, with the overall efficiency of $79 \pm 2\%$ per event, define our base sample, which  contains 1282 candidate events. The error on the efficiency includes uncertainties due to the statistics and background parameterization in the $Z \to ee$ sample. Other checks on possible variation of selection efficiency with energy, pseudorapidity, and invariant mass indicate that the efficiency is constant within uncertainties.

To  determine the  hard-scattering  vertex, we  calculate the most probable  direction of each EM object using  the transverse  and longitudinal segmentation of the EM  calorimeter~\cite{Zvvg}. Among all the reconstructed vertices in the event, we choose the one that best matches these directions. We choose this EM-object-based vertex-finding technique since it treats the diphoton and dielectron events in the same way.

We model the effects of ED via the parton-level leading-order (LO) Monte Carlo (MC) generator of Ref.~\cite{KCGL}, augmented with a parametric simulation of the D\O\ detector. The simulation takes into account detector acceptance and resolution for the EM objects, smearing and misidentification of the primary interaction vertex, initial state radiation, and the effect of different parton distributions. We used leading order CTEQ4LO~\cite{CTEQ} distributions to estimate the nominal prediction. The parameters of the detector model are tuned using independent samples of collider data. The MC includes SM contributions ($Z/\gamma^*$ and direct diphoton production), Kaluza-Klein graviton exchange diagrams, and their interference in di-EM object production.

Since the parton-level generator involves only the $2 \to 2$ hard-scattering process, we model next-to-leading order (NLO) effects by adding a transverse momentum to the di-EM system, based on the transverse momentum spectrum of di-EM events observed in the data. In the presence of the NLO corrections, the scattering angle $\theta^*$ is defined in the di-EM helicity frame, i.e., relative to the direction of the boost of the di-EM system. Since the parton-level cross section is calculated at LO, we account for NLO effects in the SM background by scaling the cross sections by a constant $K$-factor of 1.3~\cite{K-factor}. We assign a 10\% systematic uncertainty on the value of the $K$-factor to account for a possible growth of the $K$-factor at high mass. Because NLO corrections to the Kaluza-Klein diagrams have not yet been calculated, we use the same constant $K$-factor for the signal. The $K$-factor for graviton exchange is expected to grow with invariant mass, similar to that for $Z/\gamma^*$ exchange; consequently, our assumption tends to underestimate the ED contribution at high invariant mass and is conservative.

\begin{figure}[tb]
\begin{center}
\epsfxsize=3.3in
\epsffile{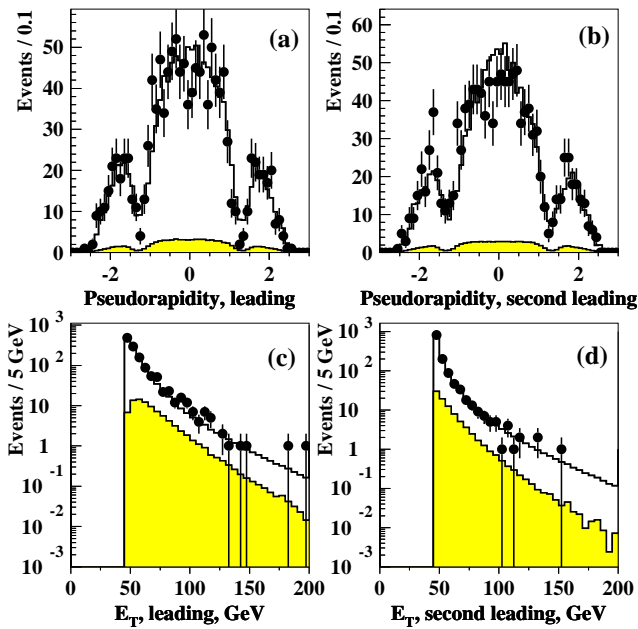} 
\vspace*{-0.05in}
\caption{
Comparison of data (points with error bars) and background predictions (solid histogram) for the pseudorapidity and transverse energy of the two EM objects in the event. The shaded histogram shows the contribution from instrumental background. The dips in pseudorapidity (defined relative to the primary interaction vertex) reflect the acceptance criteria.}
\vspace*{-0.25in}
\label{fig1}
\end{center}
\end{figure}

The main SM sources of di-EM production, Drell-Yan and direct diphoton production, are included in the MC. Other SM sources, such as $W\gamma$, $Z\gamma$, $WW$, $Z \to \tau\tau$, and $t\bar t$ production are negligible, as most have small cross section compared to that for Drell-Yan and direct diphoton processes, and are reduced further by the requirements of low $\met$ and exactly two EM objects in the event. The only non-negligible sources of background arise from single-photon and dijet events, in which jet(s) are misidentified as EM objects. We estimate this background using a separate data sample collected with a trigger requiring a single EM object. We require that a combination of a jet and the EM object satisfy all the requirements for signal, except that the EM ID requirements are applied only to the EM object. We obtain the instrumental background by scaling the number of jet-EM combinations passing our selection criteria by the probability of a jet to be mistaken as an EM object, measured to be $0.18 \pm 0.04\%$, and, within the uncertainties, is independent of $E_T$ or $\eta$~\cite{LQ}. The other source of instrumental background, $W +$ jets production with a jet misidentified as an EM object, is negligible due to the requirement of low \met. The total instrumental background in the base sample is $87 \pm 22$ events, which is less than 7\% of the dominant SM background. The sum of the SM background and the instrumental background reproduces the main kinematic characteristics of the base sample, as illustrated in Fig.~\ref{fig1}. 

The cross section in the presence of large ED is given by~\cite{GRW,HLZ,Hewett}:
\begin{eqnarray}
	\frac{d^2\sigma}{dMd\ct} & = & f_{\rm SM} + f_{\rm int}\eta_G + f_{\rm KK}\eta_G^2, \label{eq4}
\end{eqnarray}
where $f_{\rm SM}$, $f_{\rm int}$, and $f_{\rm KK}$ are functions of $(M,\ct)$ and denote the SM, interference, and $G_{\rm KK}$ terms. The effects of ED are parametrized via a single variable $\eta_G = {\cal F}/M_S^4$, where ${\cal F}$ is a dimensionless parameter of order unity, reflecting the dependence of virtual $G_{\rm KK}$ exchange on the number of extra dimensions. Different formalisms use different definitions for ${\cal F}$:
\begin{eqnarray}
	{\cal F} & = & 1, \mbox{~(GRW~\cite{GRW});} \label{eq1}\\
	{\cal F} & = & \left\{ \begin{array}{ll} 
         \log\left( \frac{M_S^2}{M} \right), & n = 2 \\
	   \frac{2}{n-2}, & n > 2
	   \end{array} \right. , \mbox{~(HLZ~\cite{HLZ});}\\ \label{eq2}
	{\cal F} & = & \frac{2\lambda}{\pi} = \pm\frac{2}{\pi}, \mbox{~(Hewett~\cite{Hewett}).} \label{eq3}
\end{eqnarray}
Note that only within the HLZ formalism ${\cal F}$ depends explicitly on $n$.

\begin{figure}[tb]
\begin{center}
\epsfxsize=3.3in
\epsffile{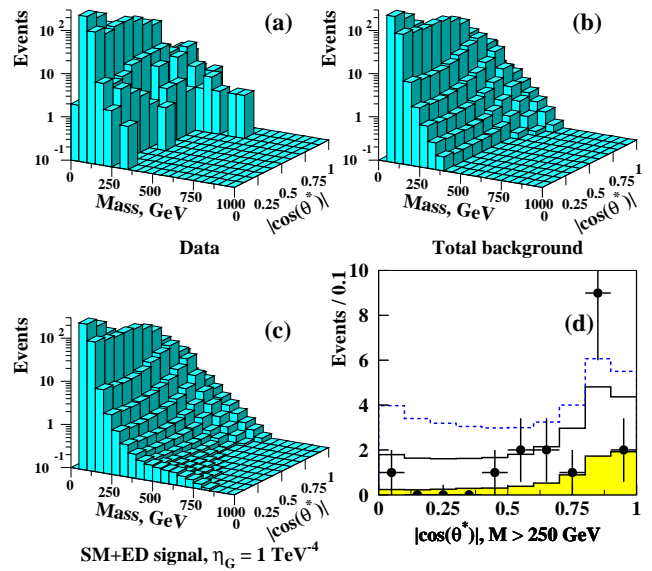} 
\vspace*{-0.1in}
\caption{
Two-dimensional distributions in di-EM mass and $|\cos\theta^*|$ for: (a) data, (b) background, (c) background and ED signal for $\eta_G = 1$~TeV$^{-4}$, and (d) $|\cos\theta^*|$ distribution for events with $M > 250$~GeV, where the filled circles correspond to the data, instrumental background is shown shaded, the entire background from SM sources is given by the solid line, and the dotted line corresponds to the sum of SM and ED for $\eta_G = 1$~TeV$^{-4}$.}
\vspace*{-0.25in}
\label{fig2}
\end{center}
\end{figure}

Figure~\ref{fig2} shows the two-dimensional distribution in $M$ and $|\ct|$ for data in (a), total background in (b), and the sum of the background and the ED signal for $\eta_G = 1$~TeV$^{-4}$ in (c). The data agree well with the background prediction and do not exhibit evidence for large ED, which would produce an excess of events at large mass and small values of $|\ct|$ (see Fig.~\ref{fig2}c). A comparison of Figs.~\ref{fig2}b and \ref{fig2}c shows that use of both $M$ and $\ct$ can improve the sensitivity over just one variable.  The projection on $\ct$ for high mass values in Fig.~\ref{fig2}d shows the relative increase of signal over SM processes at small $|\ct|$. The use of both variables allows about 10\% improvement in the sensitivity to $\eta_G$.

In the absence of evidence for ED, we set limits on the effective Planck scale. We perform a Bayesian fit of the sum of the cross section given by Eq.~(\ref{eq4}) and the instrumental background, to the data in the entire ($M$, $|\ct|$) plane shown in Fig.~\ref{fig2}, with $\eta_G$ as free parameter, with an assumed uniform prior distribution. The systematic uncertainties on signal and background in the fit include systematics of the $K$-factor (10\%), choice of parton distribution functions (5\%), integrated luminosity (4\%), EM ID efficiency (3\%), and the uncertainty on the instrumental background (25\%).

The best estimate of the parameter $\eta_G$ is consistent with the SM value of $\eta_G = 0$, and the one-sided 95\% CL limits on $\eta_G$ are:
\begin{eqnarray}
	\eta_G < & \quad\; 0.46 &~\mbox{TeV}^{-4}\; (\eta_G \ge 0) \label{eq5}\\
	\eta_G > & -0.60 &~\mbox{TeV}^{-4}\; (\eta_G \le 0), \label{eq6}
\end{eqnarray}
in good agreement with the expected sensitivity to $\eta_G$, as obtained in an ensemble of MC trials ($0.44$~TeV$^{-4}$ for $\eta_G > 0$). 
We can express these results in terms of limits on the effective Planck scale for the three formalisms of Eqs.~(\ref{eq1})--(\ref{eq3}). In the formalism of Ref.~\cite{Hewett}, both signs of $\eta_G$ are possible and therefore both limits (\ref{eq5}) and (\ref{eq6}) are relevant. In the other two formalisms, $\eta_G$ is always positive, and only the first limit is relevant. For the HLZ formalism, the case of $n = 2$ is special since ${\cal F}$, and therefore $\eta_G$, depend on $M$. To relate $\eta_G$ to $M_S$ for $n=2$, we use an average $M$ for the $G_{\rm KK}$ term at the Tevatron of (0.6~TeV)$^2$~\cite{KCGL}. Limits for different formalisms and for different numbers of extra dimensions are given in Table~\ref{table2}. They correspond to the ED compactification radius ranging from $R<0.3$~mm ($n=2$) to $R < 2$~fm ($n=7$).

\begin{table}[htb]
\vspace*{-0.1in}
\caption{Lower limits at 95\% CL on the effective Planck scale, $M_S$, in TeV.}
\label{table2}
\begin{tabular}{c|@{}cccccc|@{}cc}
\hline
GRW~\cite{GRW} & \multicolumn{6}{@{}c|}{HLZ~\cite{HLZ}} & \multicolumn{2}{@{}c}{Hewett~\cite{Hewett}} \\
\hline
& ~~$n$=2 & $n$=3 & $n$=4 & $n$=5 & $n$=6 & $n$=7~~ & ~~$\lambda=+1$ & $\lambda=-1$ \\
\cline{2-7} \cline{8-9}
1.2 & ~~1.4	& 1.4	  & 1.2   & 1.1   & 1.0   & 1.0~~  & ~~1.1      & 1.0 \\
\hline
\end{tabular}
\end{table}

In summary, we have performed the first search for large extra spatial dimensions at hadron colliders by looking for effects of virtual Kaluza-Klein gravitons in the production of dielectrons and diphotons at high energies. No evidence is found for large ED, and we set a model-independent 95\% CL upper limit of 0.46~TeV$^{-4}$ on the parameter $\eta_G$ that describes the size of the ED contribution. This result corresponds to limits on the effective Planck scale ranging between 1.0 and 1.4~TeV for several formalisms and numbers of ED. These are the most restrictive limits on large ED to date, and are complementary to analogous limits from LEP that probe a different range of invariant masses.

%
We thank the staffs at Fermilab and at collaborating institutions 
for contributions to this work, and acknowledge support from the 
Department of Energy and National Science Foundation (USA),  
Commissariat  \` a L'Energie Atomique and
CNRS/Institut National de Physique Nucl\'eaire et 
de Physique des Particules (France), 
Ministry for Science and Technology and Ministry for Atomic 
   Energy (Russia),
CAPES and CNPq (Brazil),
Departments of Atomic Energy and Science and Education (India),
Colciencias (Colombia),
CONACyT (Mexico),
Ministry of Education and KOSEF (Korea),
CONICET and UBACyT (Argentina),
A.P. Sloan Foundation,
and the A. von Humboldt Foundation.


\begin{thebibliography}{}
\bibitem{caveat}
Strictly speaking, virtual graviton effects are sensitive to the ultraviolet cutoff required to keep the divergent sum of Kaluza-Klein states finite~\protect\cite{GRW,HLZ,Hewett}. This cutoff is expected to be of the order of the effective Planck scale. Dependence on the value of the cutoff is discussed, e.g., in Ref.~\protect\cite{KCGL}.
\bibitem{eta}
Pseudorapidity $\eta_d$ is defined as $-\ln\left(\tan\frac{\theta_d}{2}\right)$, where $\theta_d$ is the polar angle with respect to the proton beam, as measured relative to the center of the detector.
\end{thebibliography}
\end{document}